\documentclass[lettersize,journal]{IEEEtran}
\usepackage[dvipdfmx]{graphicx}
\usepackage[dvipdfmx]{color}
\usepackage{amsmath,amssymb,amsfonts}
\usepackage{algorithmic}
\usepackage{algorithm}
\usepackage{array}
\usepackage[caption=false,font=normalsize,labelfont=sf,textfont=sf]{subfig}
\usepackage{textcomp}
\usepackage{stfloats}
\usepackage{url}
\usepackage{verbatim}
\usepackage{graphicx}
\usepackage{cite}
\usepackage{mathtools}
\usepackage{bm}
\usepackage{comment}
\usepackage{setspace}
\usepackage{color}
\definecolor{MYBLUE}{rgb}{0,0,1}
\newcommand{\z}[0]{ \mathcal{Z} }
\newcommand{\lss}[1]{ \!\!\mbox{ }^{#1} }
\newcommand{\rnd}[1]{ \mathrm{round}({#1}) }
\newcommand{\atantwo}[1]{ \mathrm{atan}2({#1}) }
\begin{document}

\title{Quasiperiodic Disturbance Observer\\for Wideband Harmonic Suppression}
\author{
	Hisayoshi Muramatsu
	% <-this % stops a space
	% \thanks{Manuscript received XX XXX, 20XX; revised XX XXX, 20XX.}
	% <-this % stops a space
	\thanks{H. Muramatsu is with the Mechanical Engineering Program, Hiroshima University, Higashihiroshima, Hiroshima, 739-8527, Japan (E-mail: muramatsu@hiroshima-u.ac.jp).}
	% <-this % stops a space
	\thanks{
		\color{blue}
		This paper has been published in IEEE Transactions on Control Systems Technology.\\\indent
		See https://doi.org/10.1109/TCST.2025.3566560
	}
}
\markboth{
	\color{myblue}
	This paper has been published in IEEE Transactions on Control Systems Technology.
	(See: https://doi.org/10.1109/TCST.2025.3566560)
}{}

% \IEEEpubid{0000--0000/00\$00.00~\copyright~20XX IEEE}

\maketitle
%==========================================================================
\begin{abstract}
Periodic disturbances composed of harmonics typically occur during periodic operations, impairing performance of mechanical and electrical systems.
To improve the performance, control of periodic-disturbance suppression has been studied, such as repetitive control and periodic-disturbance observers.
However, actual periodic disturbances are typically quasiperiodic owing to perturbations in each cycle, identification errors of the period, variations in the period, and/or aperiodic disturbances.
For robustness against quasiperiodicity, although wideband harmonic suppression is expected, conventional methods have trade-offs among harmonic suppression bandwidth, amplification of aperiodic disturbances, and deviation of harmonic suppression frequencies.
This paper proposes a quasiperiodic disturbance observer to compensate for quasiperiodic disturbances while simultaneously achieving the wideband harmonic suppression, non-amplification of aperiodic disturbances, and proper harmonic suppression frequencies.
A quasiperiodic disturbance is defined as comprising harmonics and surrounding signals.
On the basis of this definition, the quasiperiodic disturbance observer is designed using a periodic-pass filter of a first-order periodic/aperiodic separation filter for its Q-filter, time delay integrated with a zero-phase low-pass filter, and an inverse plant model with a first-order low-pass filter.
The periodic-pass filter achieves the wideband harmonic suppression while the zero-phase and first-order low-pass filters prevent the amplification of aperiodic disturbances and deviation of harmonic suppression frequencies.
For the implementation, the Q-filter is discretized by an exact mapping of the s-plane to the z-plane, and the inverse plant model is discretized by the backward Euler method.
The experiments validated the frequency response and position-control precision of the quasiperiodic disturbance observer while comparing it with conventional methods.
\end{abstract}


%==========================================================================
\section{Introduction}
\IEEEPARstart{P}{eriodicity} is a typical property of disturbances, which deteriorate accuracy of automatic control systems.
Periodic disturbances are caused by exogenous periodic signals and/or multiplicative modeling errors with periodic states.
For example, a periodic disturbance can result from a wind disturbance in a wind turbine \cite{2013_Houtzager_Harm}, friction force with repetitive motion in a ball-screw driven stage \cite{2014_Fujimoto_RC}, torque ripple \cite{2020_Tang_PDOB}, thrust ripple \cite{2020_Wang_Harm}, and current harmonics \cite{2019_Liu_Harm} in permanent-magnet synchronous motors, and harmonic voltage induced by nonlinear load in islanded microgrids \cite{2024_Li_Harm} and the point of common coupling for distributed generation sources \cite{2008_Patel_RC}.
Periodic disturbances comprise harmonics at integer multiples of the fundamental frequency.
Moreover, actual periodic disturbances are typically quasiperiodic, owing to perturbations in each cycle, identification errors of the period, variations in the period, and/or aperiodic disturbances.
Periodic-disturbance suppression performance against harmonics at specific frequencies easily deteriorates when disturbances are quasiperiodic.
This in turn worsens the control accuracy of automatic control systems, including the aforementioned applications.
Wideband harmonic suppression, which compensates for quasiperiodic disturbances, is expected to improve the accuracy.

Repetitive control is a classical approach for periodic-disturbance suppression \cite{1988_Hara_RC,2009_WANG_RC,2006_Bristow_RC}, which uses a time delay to acquire an internal model of the periodic disturbance \cite{2022_Mooren_RC}.
Although the exact compensation of the periodic disturbance is realized when its period is precisely known, this compensation easily deteriorates when the disturbance is quasiperiodic \cite{2002_Steinbuch_RC}.
To improve robustness against quasiperiodicity, high-order repetitive control was proposed for wideband harmonic suppression \cite{2002_Steinbuch_RC,2007_Steinbuch_RC}; however, there is a trade-off between the harmonic suppression bandwidth and aperiodic disturbance amplification.
Although there are optimal designs for repetitive control that consider this trade-off \cite{2008_PipeleersRC,2014_Chen_DOBbasedRC,2021_Nie_DOBbasedRC}, wideband harmonic suppression and non-amplification of aperiodic disturbances have not been simultaneously achieved.

Disturbance observers estimate disturbances and use them for disturbance compensation \cite{2019_Sariyildiz_DOB,2016_Chen_DOB,2015_Sariyildiz_DOB}, which do not affect the tracking performance as a two-degree-of-freedom controller.
To suppress the periodic disturbances, the sensitivity function of the disturbance observer needs to be designed.
However, higher-order design for the disturbance observer improves the sensitivity function over all low frequencies, which is an excessive improvement because harmonics exist only at specific frequencies.
Consequently, this excessive improvement worsens the sensitivity function at high frequencies owing to the Bode's sensitivity integral and robust stability owing to the trade-off between the sensitivity and complementary sensitivity functions.
According to these trade-offs, a specific disturbance observer aiming for periodic disturbances is necessary to eliminate the periodic disturbances and avoid excessive deterioration of the sensitivity function at other frequencies and robust stability.
To this end, a periodic-disturbance observer was proposed on the basis of the internal model of a periodic disturbance \cite{2018_Muramatsu_APDOB}, similar to repetitive control.
Because the periodic-disturbance observer is a two-degree-of-freedom controller, it can be applied even if the tracking command is not periodic, unlike the repetitive control.
Furthermore, there exists a combination of disturbance and periodic-disturbance observers that suppress both periodic and aperiodic disturbances \cite{2019_Muramatsu_EnPDOB}.
However, periodic-disturbance observers also have two trade-offs.
The first trade-off is between wideband harmonic suppression and deviation of harmonic suppression frequencies from harmonic frequencies \cite{2023_Tanaka_PDOB}.
Although there are designs for adjusting the first-harmonic suppression frequency to the fundamental frequency \cite{2023_Yang_PDOB} and an adaptive periodic disturbance observer for estimating the fundamental frequency of a periodic disturbance \cite{2018_Muramatsu_APDOB}, the high-order harmonic suppression frequencies still deviate.
The other trade-off is between the mitigation of harmonics and non-amplification of aperiodic disturbances \cite{2023_Yang_PDOB,2021_Lai_PDOB,2023_Li_PDOB}.

According to the aforementioned trade-offs for repetitive control and periodic-disturbance observers, no method simultaneously realizes wideband harmonic suppression, non-amplification of aperiodic disturbances, and proper harmonic suppression frequencies.
This paper proposes a quasiperiodic disturbance observer (QDOB) to solve this problem.
The scope of applicable plants is assumed to be single-input-single-output linear time-invariant plants, where zeros and poles of plant models are located on the closed left half-plane of the complex plane.
The aim of the QDOB is to compensate for quasiperiodic disturbances with the simultaneous realization of wideband harmonic suppression, non-amplification of aperiodic disturbances, and proper harmonic suppression frequencies, which is the contribution of this paper.

The QDOB is constructed on the basis of an internal model of a quasiperiodic disturbance (Section~\ref{sec:2}).
The internal model is realized by a Q-filter using a periodic-pass filter of a periodic/aperiodic separation filter \cite{2022_Muramatsu_PA,2019_Muramatsu_PASF}, where the time delay is integrated with a zero-phase low-pass filter.
The QDOB requires an inverse plant model, which is implemented with a first-order low-pass filter for stability.
These zero-phase and first-order low-pass filters realize non-amplification of aperiodic disturbances and proper harmonic suppression frequencies (Section~\ref{sec:3:ST}).
The periodic-pass filter results in wideband harmonic suppression, which is robust against quasiperiodicity (Section~\ref{sec:3:HSB}).
Nominal stability is guaranteed in Section~\ref{sec:3:ns}, and robust stability in Section~\ref{sec:3:rs} demonstrates the practicality of the proposed method with respect to the modeling errors.
To implement the QDOB, the Q-filter is discretized by an exact mapping of the s-plane to the z-plane, and the product of the inverse plant model and filter is discretized by the backward Euler method (Section~\ref{sec:4:realize}).
Experiments validated the frequency response and position-control precision of the QDOB while comparing it with conventional methods (Section~\ref{sec:5}).

\section{Quasiperiodic Disturbance Observer} \label{sec:2}
\subsection{Disturbance Observer} \label{sec:2:QD}
Consider a single-input-single-output linear time-invariant system:
\begin{align}
	\label{eq:plant}
	\mathcal{L}[y(t)]=P(s)\mathcal{L}[u(t)+v(t)],
\end{align}
with a plant $P(s)$, control input $u(t)\in \mathbb{R}$, exogenous signal $v(t)\in \mathbb{R}$, and output $y(t)\in \mathbb{R}$.
Suppose that the plant is composed of a strictly proper plant model $P_\mathrm{n}(s)$ and a modeling error $\Delta(s)$ as
\begin{subequations}
	\label{eq:Pdelta}
\begin{align}
	\label{eq:}
	P(s)&\coloneqq (1+\Delta(s))P_\mathrm{n}(s)\\
	P_\mathrm{n}(s)&\coloneqq\frac{b_{n}s^{n}+b_{n-1}s^{n-1}+\cdots+b_{1}s+b_0}{s^{m}+a_{m-1}s^{m-1}+\cdots+a_{1}s+a_0},\ n< m\\
	\Delta(s)&\coloneqq\frac{\beta_{h}s^{h}+\beta_{h-1}s^{h-1}+\cdots+\beta_{1}s+\beta_0}{s^{l}+\alpha_{l-1}s^{l-1}+\cdots+\alpha_{1}s+\alpha_0}e^{-\gamma s},\ \gamma\geq 0.
\end{align}
\end{subequations}
The roots of the numerator and denominator polynomials of $P_\mathrm{n}(s)$ are assumed to be located on the closed left half-plane of the complex plane.
In this paper, a disturbance $d(t)\in \mathbb{R}$ is defined to include both the exogenous signal and effect of the modeling error as
\begin{align}
	\label{eq:}
	\mathcal{L}[d(t)] \coloneqq \mathcal{L}[v(t)] + \Delta(s)\mathcal{L}[u(t)+v(t)],
\end{align}
and the system can be rewritten as
\begin{align}
	\label{eq:sys:yud}
	\mathcal{L}[y(t)]=P_\mathrm{n}(s)\mathcal{L}[u(t)+d(t)].
\end{align}
This system satisfies the matching condition, allowing the input $u(t)$ to compensate for the disturbance $d(t)$.

\begin{figure}[t]
	\begin{center}
		\includegraphics[width=0.95\hsize]{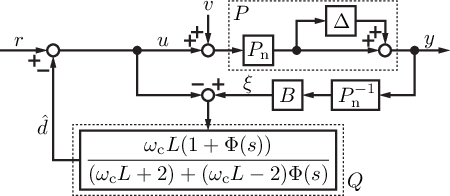}\\
		(a)\\
		\vspace{1ex}
		\includegraphics[width=0.95\hsize]{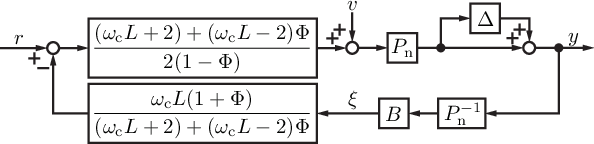}\\
		(b)
	\end{center}
	\vspace{-2ex}
	\caption{Block diagrams of the QDOB. (a) Disturbance-observer representation. (b) Equivalent single feedback-loop representation.}\label{fig:Block}
\end{figure}

To estimate the disturbance $d(t)$, the QDOB is constructed according to structure of disturbance observers as
\begin{subequations}
	\label{eq:QDOB}
\begin{align}
	\mathcal{L}[\mathcal{\xi}(t)]&= B(s)P_\mathrm{n}^{-1}(s)\mathcal{L}[y(t)]\\
	\label{eq:QDOB:Q}
	\mathcal{L}[\hat{d}(t)]&= Q(s) \mathcal{L}[\xi(t)-u(t)]\\
	u(t)&=r(t)-\hat{d}(t),
\end{align}
\end{subequations}
where $B(s)$ is set as a first-order low-pass filter
\begin{align}
	\label{eq:B}
	B(s)\coloneqq \frac{\omega_\mathrm{b}}{s+\omega_\mathrm{b}}.
\end{align}
The variables $r(t)\in \mathbb{R}$, $\hat{d}(t)\in \mathbb{R}$, and $\omega_\mathrm{b}\in \mathbb{R}_{>0}$ denote the reference signal from an outer controller, estimated disturbance, and cutoff frequency, respectively.
The filter $Q(s)$, referred to as the Q-filter, is designed on the basis of an internal model of a quasiperiodic disturbance.
A block diagram of the QDOB is presented in Fig.~\ref{fig:Block}(a).
Note that the transfer function $B(s)P_\mathrm{n}^{-1}(s)$ is a biproper or improper transfer function to be discretized using the backward Euler method.
%##############################################################
%##############################################################

%=========================================
\subsection{Quasiperiodic Disturbance} \label{sec:2:QD}
%=========================================
Let a disturbance satisfying $d(t)=d(t-L)$ be periodic with respect to a period $L\in\mathbb{R}_{>0}$.
The periodic disturbance can be expressed by the Fourier series as
\begin{align}
	\label{eq:}
	d(t)=\frac{a_0}{2}+\sum_{n=1}^{\infty}a_n\cos{\left(\frac{2n\pi}{L}t\right)}+b_n\sin{\left(\frac{2n\pi}{L}t\right)},
\end{align}
where, for a given $n$, the sum of the sine and cosine functions $a_n\cos{((2n\pi/L)t)}+b_n\sin{((2n\pi/L)t)}$ and the angular frequency $2n\pi/L$ are referred to as the $n$th harmonic and $n$th harmonic frequency, respectively.

This paper expands the periodic disturbance into a quasiperiodic disturbance based on the definition of quasiperiodicity in \cite{2022_Muramatsu_PA}.
The lifted disturbance $D_\tau(c)$ of the disturbance $d(t)$ is defined as
\begin{subequations}
	\label{eq:lifting}
\begin{align}
	D_\tau(c) &\coloneqq d(t)\ \mathrm{s.t.}\ t= cL+\tau\\
	D&:\{\tau\in \mathbb{T}\} \times \{c\in \mathbb{Z}\} \to \mathbb{R},
\end{align}
\end{subequations}
where $\mathbb{T}\coloneqq [0,L)$.
The arguments $c$ and $\tau$ denote the cycle and elapsed time within the cycle, respectively.
Note that the lifted periodic disturbance satisfies $D_\tau(c)=D_\tau(c-1)$ for all cycles.
Subsequently, let $\tilde{\mathcal{D}}:\mathbb{T}\times\mathbb{R}\to \mathbb{C}$ such that
\begin{subequations}
	\label{eq:Fourier:D}
\begin{align}
	\mathcal{F}[D_\tau(c)]&=\tilde{\mathcal{D}}_\tau(\omega)
	\lor
	D_\tau(c)=\mathcal{F}^{-1}[\tilde{\mathcal{D}}_\tau(\omega)]\\
	\mathcal{F}[D_\tau(c)]&\coloneqq \sum_{c=-\infty}^\infty D_\tau(c) e^{-j\omega Lc}\\
	\mathcal{F}^{-1}[\tilde{\mathcal{D}}_\tau(\omega)]&\coloneqq \frac{L}{2\pi}\int_{-\pi/L}^{\pi/L}\tilde{\mathcal{D}}_\tau(\omega)e^{j\omega Lc}d\omega
\end{align}
\end{subequations}
be the discrete-time Fourier-transform lifted disturbance, where $\mathcal{F}$ and $\mathcal{F}^{-1}$ denote the discrete-time Fourier and discrete-time inverse Fourier transforms, respectively.
$\lor$ stands for the logical disjunction.
In this paper, the function $\tilde{\mathcal{D}}$ that satisfies
\begin{align}
	\label{eq:def:quasiperiodicity}
	\tilde{\mathcal{D}}_\tau(\omega)=0,\ \forall&\omega\in\{\omega\in\mathbb{R}||\omega|>\rho\},\ \forall \tau\in\mathbb{T}
\end{align}
is called quasiperiodic, where $\omega\in \mathbb{R}$ is the angular frequency and $\rho\in[0,\pi/L)$ is referred to as the separation frequency.
Using these, a set of functions of quasiperiodic disturbances is defined as
\begin{align}
	\label{eq:}
	\mathbb{P}_\rho\coloneqq \{d:\mathbb{R}\to\mathbb{R}|&\eqref{eq:lifting}\land \eqref{eq:Fourier:D}\land \eqref{eq:def:quasiperiodicity}\},
\end{align}
where $\land$ denotes the logical conjunction.
Consequently, the disturbance $d(t)$ is called quasiperiodic with respect to the separation frequency $\rho$ if $d\in\mathbb{P}_\rho$, which is the definition of a quasiperiodic disturbance in this paper.
This definition implies that the quasiperiodic disturbance has low-frequency changes or zero and does not exhibit high-frequency changes over cycles $c$.
The boundary between the low  and high frequencies is the separation frequency $\rho$.
The separation frequency $\rho$ is the degree of the quasiperiodicity, and the quasiperiodic disturbance with respect to $\rho=0$ rad/s is equivalent to the periodic disturbance such that $d(t)=d(t-L)$.

\subsection{Q-Filter}
The Q-filter in \eqref{eq:QDOB:Q} is designed to estimate the quasiperiodic disturbance $d(t)$ by realizing the internal model of the quasiperiodic disturbance as a periodic-pass filter of a first-order periodic/aperiodic separation filter proposed in \cite{2019_Muramatsu_PASF,2022_Muramatsu_PA}.
Because the lifted quasiperiodic disturbance $D_\tau(c)$ satisfies \eqref{eq:def:quasiperiodicity}, it comprises low-frequency signals at frequencies less than or equal to the separation frequency $\rho$.
Hence, a first-order low-pass filter can extract the lifted quasiperiodic disturbance $D_\tau(c)$ from the lifted error $\Xi_\tau(c)-U_\tau(c)$ approximately as
\begin{align}
	\label{eq:def:zlifted:Qfilter}
	\frac{\mathcal{Z}[{D}_\tau(c)]}{\mathcal{Z}[\Xi_\tau(c)-U_\tau(c)]}\approx\frac{\omega_\mathrm{c} L(1+Z^{-1})}{(\omega_\mathrm{c} L+2)+(\omega_\mathrm{c} L-2)Z^{-1}},
\end{align}
where $\Xi_\tau(c)$ and $U_\tau(c)$ are the lifted functions of $\xi(t)$ and $u(t)$ in \eqref{eq:QDOB}, respectively.
Note that the z-transform with respect to $Z$ is based on the cycle $c$ with the sampling time $L$, which is the cycle period.
The cutoff frequency $\omega_\mathrm{c}$ is designed in Section~\ref{sec:3:HSB}.
The z-domain low-pass filter for discrete-time lifted signals with respect to $c$ is transformed into an s-domain filter using the exact mapping $Z^{-1}=e^{-Ls}$ as
\begin{align}
	\label{eq:periodic-pass}
	\frac{\mathcal{L}[{d}(t)]}{\mathcal{L}[\xi(t)-u(t)]}&\approx\frac{\omega_\mathrm{c} L(1+e^{-Ls})}{(\omega_\mathrm{c} L+2)+(\omega_\mathrm{c} L-2)e^{-Ls}},
\end{align}
which is the periodic-pass filter of the first-order periodic/aperiodic separation filter.

\begin{figure}[t!]
	\begin{center}
		\includegraphics[width=0.9\hsize]{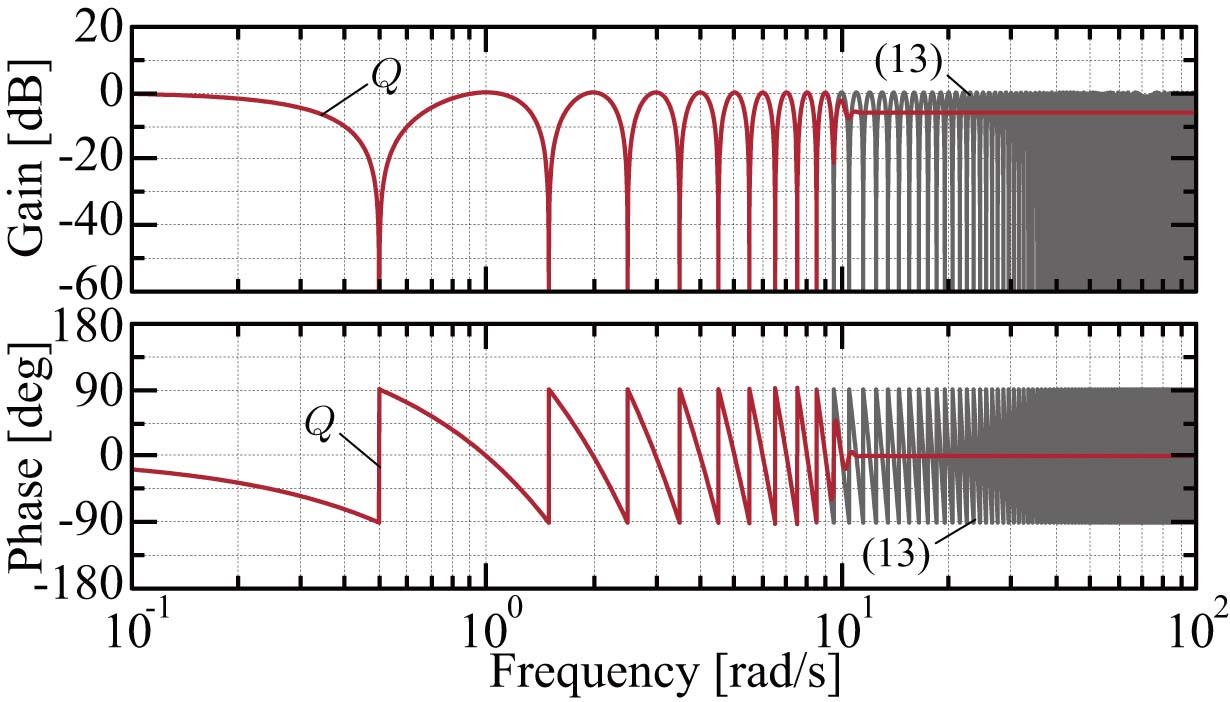}
	\end{center}
	\vspace{-2ex}
	\caption{Bode plot of the Q-filter \eqref{eq:Q-filter} of the QDOB and the first-order periodic-pass filter in \eqref{eq:periodic-pass}. The parameters are $l=3$, $N_\mathrm{max}=256$, $\omega_\mathrm{a}=10\ \mathrm{rad/s}$, $\omega_\mathrm{c}=2/L$, $L=2\pi\ \mathrm{s}$, and $T=1.0\times10^{-4}\ \mathrm{s}$.}\label{fig:Q}
\end{figure}

Although the time delays $e^{-Ls}$ of \eqref{eq:periodic-pass} are necessary for quasiperiodic disturbance suppression, they induce amplification of aperiodic disturbances and deviation of harmonic suppression frequencies.
Thus, a zero-phase low-pass filter is combined with each time delay to limit the frequencies at which the time delay affects.
Consequently, the Q-filter of the QDOB is set to
\begin{align}
	\label{eq:Q-filter}
	Q(s)&\coloneqq\frac{\omega_\mathrm{c} L(1+\Phi(s))}{(\omega_\mathrm{c} L+2)+(\omega_\mathrm{c} L-2)\Phi(s)},
\end{align}
where $\Phi(s)$ is the linear-phase low-pass filter, which is the product of the time delay and zero-phase low-pass filter.
The equivalent block diagram using the Q-filter in Fig.~\ref{fig:Block}(b) shows that the controller has the denominator $2(1-\Phi(s))$, which is the disturbance generating polynomial for the internal model principle \cite{goodwin2001control}.
Fig.~\ref{fig:Q} depicts the Bode plot of the Q-filter \eqref{eq:Q-filter} and first-order periodic-pass filter \eqref{eq:periodic-pass} with representative parameters, which shows that the effect of the time delay on the Q-filter is mitigated from around the cutoff frequency $\omega_\mathrm{a}=10\ \mathrm{rad/s}$.
%##############################################################
%##############################################################

\subsection{Linear-Phase Low-Pass Filter} \label{sec:LP-LPF}
\begin{figure}[t!]
	\begin{center}
		\includegraphics[width=0.9\hsize]{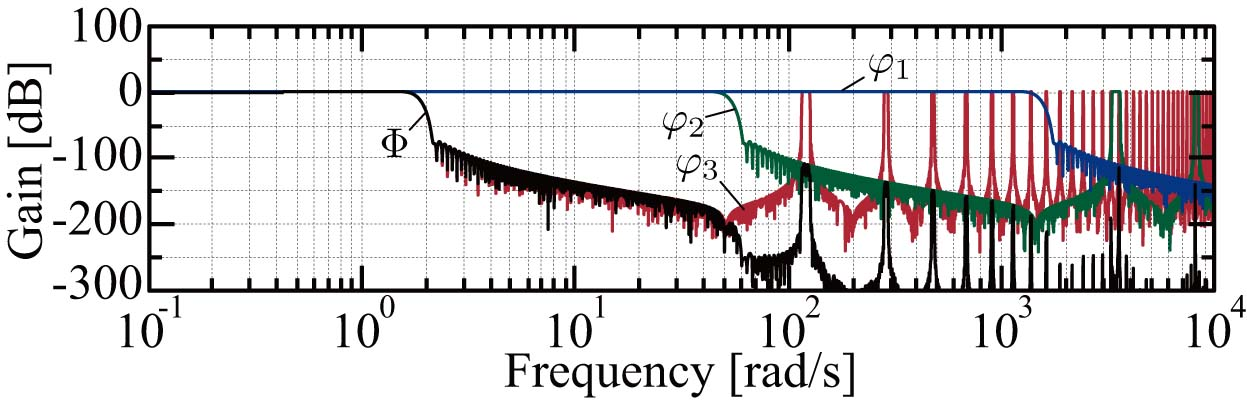}
	\end{center}
	\vspace{-2ex}
	\caption{Gains of the three-stage linear-phase low-pass filter $\Phi(s)$ and the filters $\varphi_1$, $\varphi_2$, and $\varphi_3$ for each stage. The parameters are $l=3$, $N_\mathrm{max}=256$, $\omega_\mathrm{a}=10\ \mathrm{rad/s}$, $L=2\pi\ \mathrm{s}$, and $T=1.0\times10^{-4}\ \mathrm{s}$.}\label{fig:MultiFIR}
\end{figure}

This section provides a design example of the linear-phase low-pass filter $\Phi(s)$ in the Q-filter \eqref{eq:Q-filter}.
Prior to the design, $\Phi(s)$ is divided into a sample delay $e^{-Ts}$ with sampling time $T\in \mathbb{R}_{>0}$ and a multistage linear-phase low-pass filter $\Psi(s)$ for implementation in discrete time
\begin{align}
	\label{eq:}
	\Phi(s)=e^{-Ts}\Psi(s),\
	\Psi(s)\coloneqq e^{(T-L)s}\prod_{i=1}^{l}H_i(s),
\end{align}
where $H_i(s)$ is the $i$th-stage zero-phase low-pass filter and $l\in \mathbb{Z}_{>0}$ denotes the number of stages.
A multistage design is employed to realize a low cutoff frequency with less computational cost.
The filter for each stage $H_i(s)$ is defined as
\begin{subequations}
	\label{eq:}
\begin{align}
	&H_i(s)\coloneqq \frac{\sum_{n=-\infty}^{\infty}w(n,N)h(n,\omega_i,U_i)e^{nU_is}}{\sum_{n=-\infty}^{\infty}w(n,N)h(n,\omega_i,U_i)}\\
	&h(n,\omega_i,U_i)\coloneqq\left\{
	\begin{array}{cl}
		{U_i\omega_i}/{\pi}&\mathrm{if}\ n=0\\
		{\sin(nU_i\omega_i)}/{(n\pi)}&\mathrm{if}\ n\neq0\\
	\end{array}
	\right.\\
	&w(n,N)\coloneqq\left\{
	\begin{array}{cl}
		\multicolumn{2}{l}{0.42+0.5\cos({n\pi}{/N})}\\
		+0.08\cos({2n\pi}/{N})&\mathrm{if}\ |n|\leq N\\
		0&\mathrm{if}\ |n|>N,
	\end{array}
	\right.
\end{align}
\end{subequations}
where the coefficient $h(n,\omega_i,U_i)\in \mathbb{R}$ can be derived by the inverse Fourier transform of the frequency characteristic of an ideal zero-phase low-pass filter, and the Blackman window $w(n,N)\in \mathbb{R}$ extracts a finite number of coefficients.
By multiplying $e^{(T-L)s}$ and $\prod_{i=1}^{l}H_i(s)$, the filter $\Psi(s)$ becomes
\begin{subequations}
	\label{eq:}
\begin{align}
	\Psi(s)&=e^{(T-L+N\sum_{i=1}^lU_i)s}\prod_{i=1}^{l}\varphi_i(s)\\
	\varphi_i(s) &\coloneqq\frac{\sum_{n=-N}^{N}w(n,N)h(n,\omega_i,U_i)e^{(n-N)U_is}}{\sum_{n=-N}^{N}w(n,N)h(n,\omega_i,U_i)},
\end{align}
\end{subequations}
where the $i$th-stage sampling time $U_i\in \mathbb{R}_{>0}$, $i$th-stage cutoff frequency $\omega_i\in \mathbb{R}_{>0}$, and order $N\in \mathbb{R}_{>0}$ are determined as follows
\begin{subequations}
	\label{eq:}
\begin{align}
	\label{eq:Ui}
	U_{i}&\coloneqq\left\{
	\begin{array}{cl}
		T&\mathrm{if}\ i=1\\
		{\pi}/{\omega_{i-1}}&\mathrm{otherwise}
	\end{array}
	\right.\\
	\label{eq:wi}
	\omega_i&\coloneqq\left\{
	\begin{array}{cl}
		\omega_\mathrm{a}&\mathrm{if}\ i=l\\
		{2c\pi}/{U_i}&\mathrm{otherwise},
	\end{array}
	\right.\
	c=\frac{1}{2}\left(\frac{T\omega_\mathrm{a}}{\pi}\right)^{1/l}\\
	N&\coloneqq\min\{\max(\mathcal{N}),\ N_\mathrm{max}\}\\
	\mathcal{N} &\coloneqq \{n\in \mathbb{Z}_{>0}| n \leq (L-T)/\textstyle\sum_{j=1}^iU_j\}.
\end{align}
\end{subequations}
The sampling time $U_1$ is set to the original sampling time $T$ for the first stage.
Meanwhile, it is set for the other stages so that the $i$th-stage Nyquist frequency $\pi/U_{i}$ equals the cutoff frequency of the previous stage $\omega_{i-1}$.
The cutoff frequency $\omega_i$ decreases over the stages, and the coefficient $c$ is derived such that $U_1=T$ and $\omega_l=\omega_\mathrm{a}$.
The order $N$ is maximized in the set $\mathcal{N}$ of orders that make the filter $\Psi(s)$ causal without exceeding the maximum order $N_\mathrm{max}$ given by allowed computational cost.
Fig.~\ref{fig:MultiFIR} depicts the gain of the three-stage linear-phase low-pass filter $\Phi(s)$ with representative parameters, where the stage-by-stage reduction in the cutoff frequency can be observed.
%##############################################################
%##############################################################

%##############################################################
\begin{figure}[t!]
	\begin{center}
		\includegraphics[width=0.9\hsize]{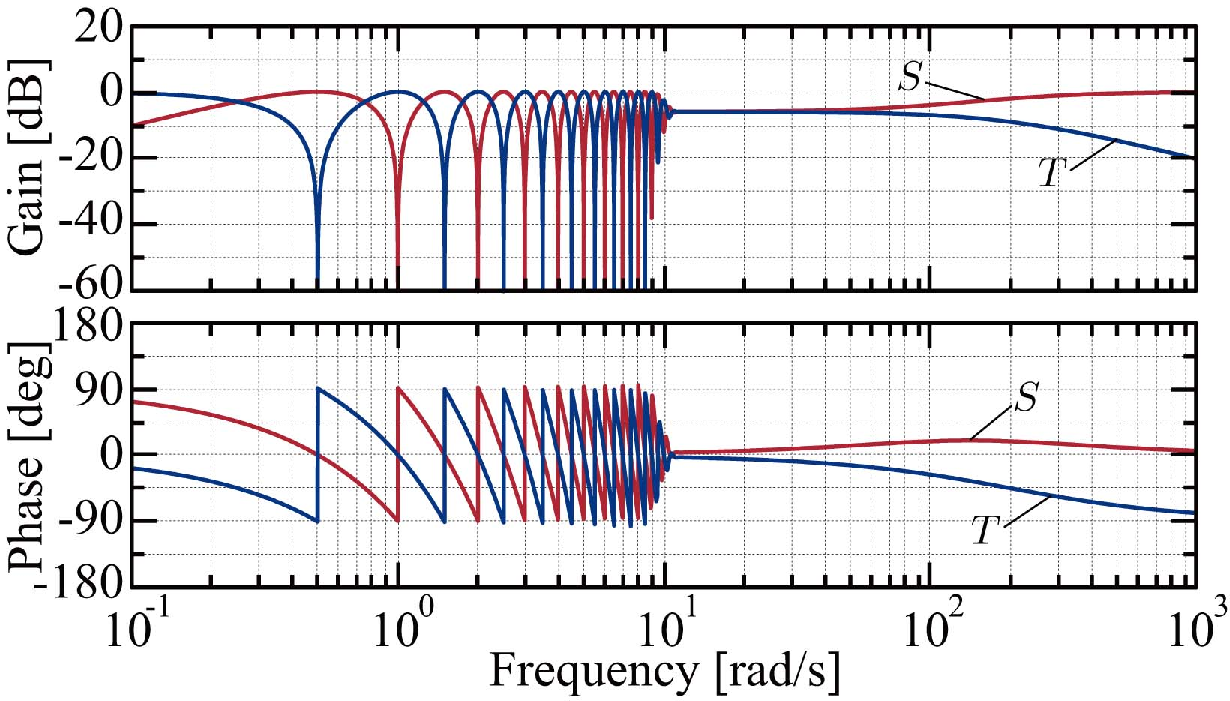}\\
		(a)\\
		\vspace{1ex}
		\includegraphics[width=0.9\hsize]{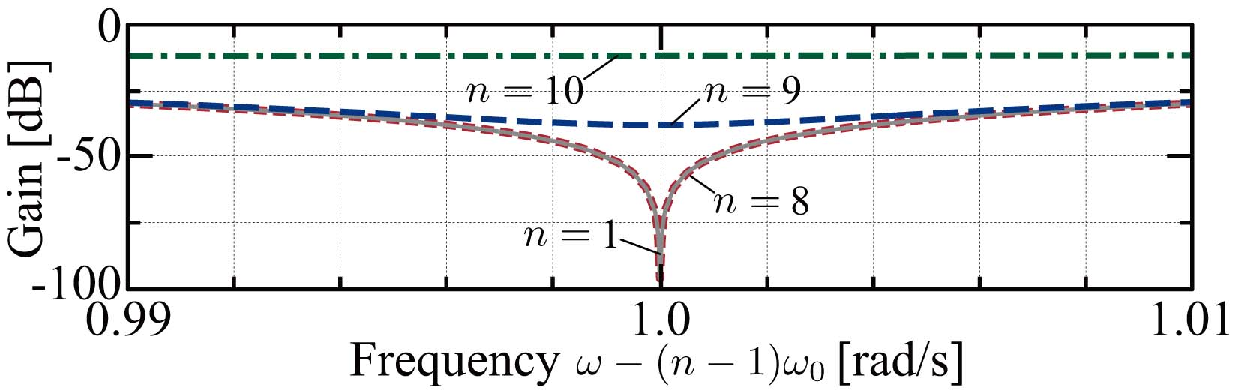}\\
		(b)
	\end{center}
	\vspace{-2ex}
	\caption{Bode plots of the sensitivity and complementary sensitivity functions in \eqref{eq:SandT}. The parameters are $l=3$, $N_\mathrm{max}=256$, $\omega_\mathrm{a}=10\ \mathrm{rad/s}$, $\omega_{\mathrm{b}}=100\ \mathrm{rad/s}$, $\omega_\mathrm{c}=2/L$, $L=2\pi/\omega_0$, $\omega_0=1\ \mathrm{rad/s}$, and $T=1.0\times10^{-4}\ \mathrm{s}$. (a) Sensitivity and complementary sensitivity functions. (b) Enlarged view of the gain of the sensitivity function at the $n$th harmonic frequencies.}\label{fig:Block:ST}
\end{figure}
%##############################################################
\section{Design and Analysis} \label{sec:3}
\subsection{Sensitivity and Complementary Sensitivity Functions}\label{sec:3:ST}
The cutoff frequencies $\omega_\mathrm{a}$ in \eqref{eq:wi} and $\omega_\mathrm{b}$ in \eqref{eq:B} are designed according to the sensitivity function (index for disturbance suppression) and complementary sensitivity function (index for robust stability and noise sensitivity).
Suppose no modeling error $\Delta=0$.
Then, the open-loop transfer function $\Gamma(s)$ is
\begin{align}
	\label{eq:Gamma}
	\Gamma(s)&=\frac{\omega_\mathrm{c} L}{2}\frac{1+\Phi(s)}{1-\Phi(s)}B(s)=\frac{\omega_\mathrm{c} L}{2}\frac{1+\Phi(s)}{1-\Phi(s)}\frac{\omega_\mathrm{b}}{s+\omega_\mathrm{b}}
\end{align}
according to Fig.~\ref{fig:Block}(b).
Using the open-loop transfer function, the sensitivity function $S(s)$ and complementary sensitivity function $T(s)$ are defined and calculated as
\begin{subequations}
	\label{eq:SandT}
\begin{align}
	\label{eq:S}
	S(s)&\coloneqq \frac{1}{1+\Gamma}
	=\frac{2(1-\Phi)}{(\omega_\mathrm{c} LB+2)+(\omega_\mathrm{c} LB-2)\Phi}\\
	\label{eq:T}
	T(s)&\coloneqq \frac{\Gamma}{1+\Gamma}
	=\frac{\omega_\mathrm{c} L(1+\Phi)B}{(\omega_\mathrm{c} LB+2)+(\omega_\mathrm{c} LB-2)\Phi}.
\end{align}
\end{subequations}
The sensitivity function satisfies $\mathcal{L}[y]/\mathcal{L}[d]=P_\mathrm{n}(s)S(s)$.

The sensitivity and complementary sensitivity functions show different features in the three frequency ranges: $0\leq\omega\ll\omega_\mathrm{a}$, $\omega_\mathrm{a}\ll\omega\ll\omega_{\mathrm{b}}$, and $\omega_{\mathrm{b}}\ll\omega\ll\pi/T$, as shown in Fig.~\ref{fig:Block:ST}(a).
Note that $\pi/T$ corresponds to the Nyquist frequency.
For each range, the transfer functions can be approximated as follows
\begin{subequations}
	\label{eq:ST:freqranges}
\begin{align}
	&\hspace{-1em}0\leq\omega\ll\omega_\mathrm{a}:\notag\\
	\label{eq:S:lower}
	&S(j\omega)\approx\frac{2(1-e^{-jL\omega})}{(\omega_\mathrm{c} L+2)+(\omega_\mathrm{c} L-2)e^{-jL\omega}}\\
	\label{eq:T:lower}
	&T(j\omega)\approx\frac{\omega_\mathrm{c} L(1+e^{-jL\omega})}{(\omega_\mathrm{c} L+2)+(\omega_\mathrm{c} L-2)e^{-jL\omega}}\\
	&\hspace{-1em}\omega_\mathrm{a}\ll\omega\ll \omega_\mathrm{b}:\notag\\
	\label{eq:ST:middle}
	&S(j\omega)\approx\frac{2}{\omega_\mathrm{c} L+2},\
	T(j\omega)\approx\frac{\omega_\mathrm{c} L}{\omega_\mathrm{c} L+2}\\
	&\hspace{-1em}\omega_{\mathrm{b}}\ll\omega\ll\pi/T:\notag\\
	\label{eq:ST:higher}
	&S(j\omega)\approx \frac{2}{\omega_\mathrm{c} LB(j\omega)+2},\
	T(j\omega)\approx \frac{\omega_\mathrm{c} LB(j\omega)}{\omega_\mathrm{c} LB(j\omega)+2},
\end{align}
\end{subequations}
which are based on the approximations of the low-pass filters: $|\Phi(j\omega)|\approx1$ if $\omega\ll\omega_\mathrm{a}$, $\Phi(j\omega)\approx0$ if $\omega_\mathrm{a}\ll\omega\ll\pi/T$, and $B(j\omega)\approx1$ if $\omega\ll\omega_\mathrm{b}$.
The lower-frequency range $\omega\ll\omega_\mathrm{a}$ is for quasiperiodic disturbance suppression, where the sensitivity function $S(s)$ approximately becomes the periodic-pass filter \eqref{eq:periodic-pass}.
In the higher-frequency range $\omega_{\mathrm{b}}\ll\omega\ll\pi/T$, the low-pass filter $B(s)$ is dominant for highly robust stability and low noise sensitivity via the complementary sensitivity function $T(s)$.
Lastly, the middle-frequency range $\omega_\mathrm{a}\ll\omega\ll\omega_{\mathrm{b}}$, which separates the higher- and lower-frequency ranges, rejects amplification of aperiodic disturbances and deviation of the harmonic suppression frequencies, as shown in Fig.~\ref{fig:Block:ST}(a) and Fig.~\ref{fig:Block:ST}(b), respectively.

\subsection{Harmonic Suppression Bandwidth}\label{sec:3:HSB}
\begin{figure}[t!]
	\begin{minipage}{0.49\hsize}
		\begin{center}
			\includegraphics[width=0.95\hsize]{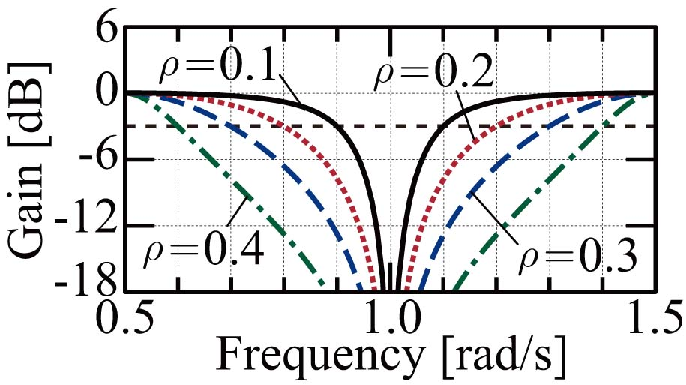}
		\end{center}
	\end{minipage}
	\begin{minipage}{0.49\hsize}
		\begin{center}
			\includegraphics[width=0.95\hsize]{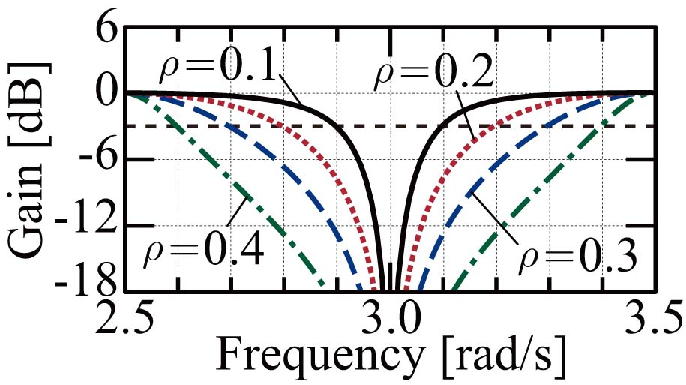}
		\end{center}
	\end{minipage}\\ \\
	\begin{minipage}{0.49\hsize}
		\begin{center}
			(a) First harmonic.
		\end{center}
	\end{minipage}
	\begin{minipage}{0.49\hsize}
		\begin{center}
			(b) Third harmonic.
		\end{center}
	\end{minipage}\\ \\
	\begin{minipage}{0.49\hsize}
		\begin{center}
			\includegraphics[width=0.95\hsize]{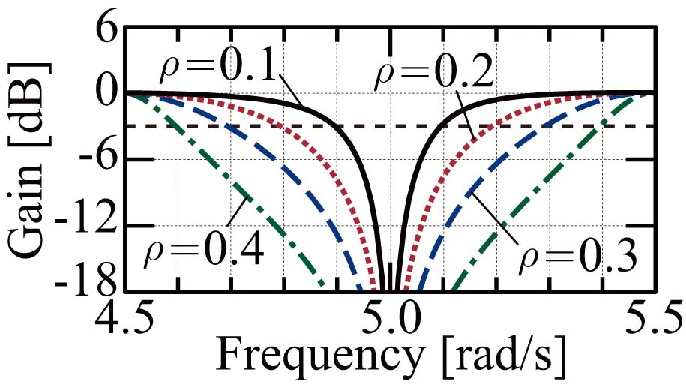}
		\end{center}
	\end{minipage}
	\begin{minipage}{0.49\hsize}
		\begin{center}
			\includegraphics[width=0.95\hsize]{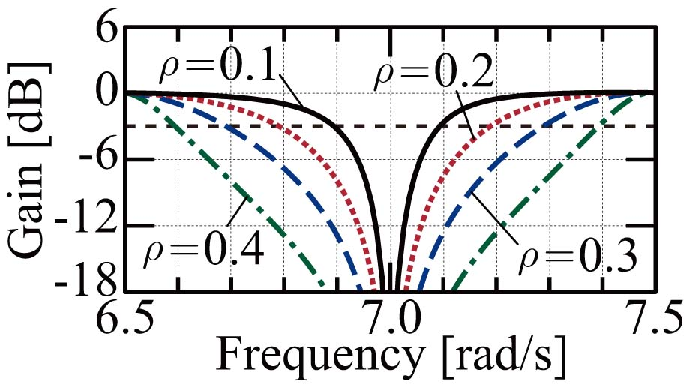}
		\end{center}
	\end{minipage}\\ \\
	\begin{minipage}{0.49\hsize}
		\begin{center}
			(c) Fifth harmonic.
		\end{center}
	\end{minipage}
	\begin{minipage}{0.49\hsize}
		\begin{center}
			(d) Seventh harmonic.
		\end{center}
	\end{minipage}
	% \vspace{-2ex}
	\caption{Bode plots of the sensitivity function in \eqref{eq:S} with the design of the cutoff frequency $\omega_\mathrm{c}$ in \eqref{eq:rho} and various separation frequencies $\rho$.
	The parameters are $l=3$, $N_\mathrm{max}=256$, $\omega_\mathrm{a}=10\ \mathrm{rad/s}$, $\omega_{\mathrm{b}}=100\ \mathrm{rad/s}$, $L=2\pi/\omega_0$, $\omega_0=1\ \mathrm{rad/s}$, and $T=1.0\times10^{-4}\ \mathrm{s}$.}\label{fig:wideband}
\end{figure}

The suppression bandwidth around the harmonic frequencies is designed via the cutoff frequency $\omega_\mathrm{c}$ of the Q-filter \eqref{eq:Q-filter}.
In the frequency range $0\leq\omega\ll\omega_\mathrm{a}$ for quasiperiodic disturbance suppression, the gain of the approximate sensitivity function \eqref{eq:S:lower} is
\begin{align}
	\label{eq:}
	\left|\frac{2(1-e^{-Ls})}{(\omega_\mathrm{c} L+2)+(\omega_\mathrm{c} L-2)e^{-Ls}}\right|=\sqrt{\frac{4\tan^2(L\omega/2)}{\omega_\mathrm{c}^2L^2+4\tan^2(L\omega/2)}}.
\end{align}
By determining the cutoff frequency $\omega_\mathrm{c}$ using the separation frequency $\rho$ as
\begin{align}
	\label{eq:rho}
	\omega_\mathrm{c}\coloneqq\frac{2}{L}\tan{\left(\frac{L}{2}\rho\right)},
\end{align}
the gain of the sensitivity function satisfies
\begin{align}
	\label{eq:-3db:wc}
	20\log|S(j(n\omega_0\pm\rho))|\approx -3\ \mathrm{dB}.
\end{align}
The harmonic suppression bandwidth ranges from $n\omega_0-\rho$ to $n\omega_0+\rho$ around a harmonic frequency $n\omega_0$, in which the gain is less than $-3\ \mathrm{dB}$.
Hence, an increase in the cutoff frequency $\rho$ achieves wideband harmonic suppression.
Fig.~\ref{fig:wideband} shows that \eqref{eq:-3db:wc} holds for various separation frequencies from the first to the seventh harmonic frequencies under the representative parameters.
%##############################################################
%##############################################################
%##############################################################
\begin{figure}[t!]
	\begin{center}
		\includegraphics[width=0.9\hsize]{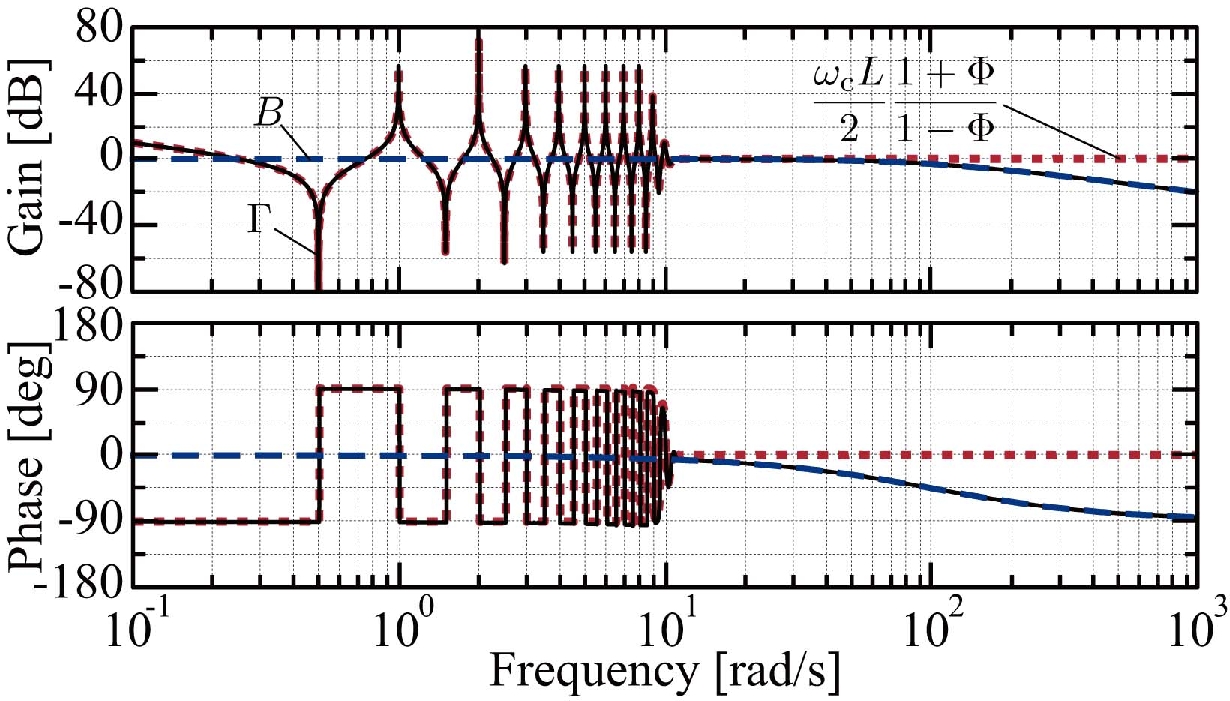}
	\end{center}
	\vspace{-2ex}
	\caption{Bode plot of the open-loop transfer function.
	The parameters are $l=3$, $N_\mathrm{max}=256$, $\omega_\mathrm{a}=10\ \mathrm{rad/s}$, $\omega_{\mathrm{b}}=100\ \mathrm{rad/s}$, $\omega_\mathrm{c}=2/L$, $L=2\pi/\omega_0$, $\omega_0=1\ \mathrm{rad/s}$, and $T=1.0\times10^{-4}\ \mathrm{s}$.}\label{fig:Block:OL}
\end{figure}
%##############################################################

\subsection{Nominal Stability}\label{sec:3:ns}
Consider nominal stability with an asymptotically stable plant model $P_\mathrm{n}(s)$ whose poles are in the open left half-plane in the complex plane.
In the open-loop transfer function \eqref{eq:Gamma}, the phase of the first-order low-pass filter $B(s)$ satisfies
\begin{align}
	\label{eq:angle:B:ranges}
	-\pi/2<\angle B(j\omega)=\atantwo{-\omega,\ \omega_\mathrm{b}}\leq 0,\ \forall\omega\in \mathbb{R}_{\geq0}.
\end{align}
Using the linear-phase characteristic $\Phi(j\omega)=|\Phi(j\omega)|e^{-jL\omega}$ of $\Phi(s)$, the phase of the other part can be calculated as
\begin{align}
	\label{eq:angle:phi}
	&\angle\frac{\omega_\mathrm{c} L}{2}\frac{1+|\Phi(j\omega)|e^{-jL\omega}}{1-|\Phi(j\omega)|e^{-jL\omega}}\notag\\
	&\hspace{1em}=\angle\frac{\omega_\mathrm{c} L}{2}\dfrac{(1 - |\Phi(j\omega)|^2) - j2|\Phi(j\omega)|\sin(L\omega)}{1 - 2 |\Phi(j\omega)| \cos(L\omega) + |\Phi(j\omega)|^2}\notag\\
	&\hspace{1em}=\atantwo{- 2|\Phi(j\omega)|\sin(L\omega),\ 1 - |\Phi(j\omega)|^2}.
\end{align}
Assume that the gain of the linear-phase low-pass filter satisfies $|\Phi(j\omega)|\leq1$, which is practical according to Fig.~\ref{fig:MultiFIR}.
Then, $1 - |\Phi(j\omega)|^2$ is non-negative, and the phase satisfies
\begin{align}
	\label{eq:angle:phi:ranges}
	-\frac{\pi}{2}\leq\angle\left(\frac{\omega_\mathrm{c} L}{2}\dfrac{1+\Phi(j\omega)}{1-\Phi(j\omega)}\right)\leq\frac{\pi}{2},\ \forall\omega\in \mathbb{R}_{\geq0}.
\end{align}
Based on \eqref{eq:angle:B:ranges} and \eqref{eq:angle:phi:ranges}, the overall phase of the open-loop transfer function \eqref{eq:Gamma} exists within
\begin{align}
	-\pi<\angle\Gamma(j\omega)\leq\frac{\pi}{2},\ \forall\omega\in \mathbb{R}_{\geq0}
\end{align}
and does not reach $-\pi$ rad/s.
On the basis of the Nyquist stability criterion, the system is nominally stable.

The stability margin is extended by designing the cutoff frequencies $\omega_\mathrm{a}$ and $\omega_\mathrm{b}$ such that $\omega_\mathrm{a}\ll\omega_\mathrm{b}$, as described in Section~\ref{sec:3:ST}.
By satisfying $\omega_\mathrm{a}\ll\omega_\mathrm{b}$, the phase range narrows to approximately $-\pi/2\leq\angle\Gamma(j\omega)\leq\pi/2$, and a phase margin of $-\pi/2$ rad is secured within a frequency range less than the Nyquist frequency, as shown in the phase plot of Fig.~\ref{fig:Block:OL}.
Additionally, the low-pass filter $B(s)$ extends the gain margin at frequencies greater than $\omega_\mathrm{b}$, as shown in the gain plot of Fig.~\ref{fig:Block:OL}.

\subsection{Robust Stability}\label{sec:3:rs}
Consider robust stability against a stable modeling error $\Delta(s)$.
Let $\tilde{T}(s)$ be
\begin{align}
	\label{eq:}
	\tilde{T}(s)\coloneqq\left\{
	\begin{array}{cl}
		\dfrac{\omega_\mathrm{c} L(1+e^{-Ls})}{(\omega_\mathrm{c} L+2)+(\omega_\mathrm{c} L-2)e^{-Ls}}&\mathrm{if}\ \omega\leq \omega_\mathrm{a}\\
		\dfrac{\omega_\mathrm{c} LB(s)}{\omega_\mathrm{c} LB(s)+2}&\mathrm{if}\ \omega>\omega_\mathrm{a},
	\end{array}
	\right.
\end{align}
which is the right-hand side of \eqref{eq:ST:freqranges}.
Subsequently, the complementary sensitivity function $T$ is decomposed into $T(s)=\sigma(s)\tilde{T}(s)$, where $\sigma(s)$ is the approximation error of \eqref{eq:ST:freqranges}.
The approximation error $\sigma(s)$ decreases as $\omega_\mathrm{b}$ increases relative to $\omega_\mathrm{a}$.
Assume that the worst error $\tilde{\Delta}(s)$ against both the modeling error $\Delta(j\omega)$ and approximation error $\sigma(j\omega)$ is known, where $|\sigma(j\omega)\Delta(j\omega)|<\tilde{\Delta}(\omega),\ \forall\omega\in\mathbb{R}_{\geq0}$.

Suppose the system is nominally stable on the basis of Section~\ref{sec:3:ns}.
Then, the robust stability condition based on the small gain theorem is
\begin{align}
	\label{eq:robust:condition}
	|\tilde{T}(j\omega)|\leq\tilde{\Delta}^{-1}(\omega),\ \forall\omega\in\mathbb{R}_{\geq0}.
\end{align}
The gain of $\tilde{T}$ can be calculated as
\begin{subequations}
	\label{eq:}
\begin{align}
	\label{eq:Ttilde:a}
	|\tilde{T}(j\omega)|
	&=\frac{\omega_\mathrm{c} L}{\sqrt{\omega_\mathrm{c}^2L^2+4\tan^2(L\omega/2)}},\ \mathrm{if}\ \omega\leq \omega_\mathrm{a}\\
	\label{eq:Ttilde:b}
	|\tilde{T}(j\omega)|
	&=\frac{\omega_\mathrm{c} L \omega_\mathrm{b}}{\sqrt{4 \omega^2 + (2 + \omega_\mathrm{c} L)^2 \omega_\mathrm{b}^2}},\ \mathrm{if}\ \omega>\omega_\mathrm{a},
\end{align}
\end{subequations}
where the gain \eqref{eq:Ttilde:a} in $\omega\leq \omega_\mathrm{a}$ is less than or equal to 1, and the gain \eqref{eq:Ttilde:b} in $\omega>\omega_\mathrm{a}$ decreases as $\omega_\mathrm{b}$ and/or $\omega_\mathrm{c}$ decreases.
The cutoff frequency $\omega_\mathrm{c}$ is determined by the separation frequency $\rho$ as \eqref{eq:rho}.
Consequently, the angular frequencies: $\omega_\mathrm{a}$, $\omega_\mathrm{b}$, and $\rho$ need to be low enough to satisfy the condition \eqref{eq:robust:condition}.

\section{Realization} \label{sec:4}
\subsection{Discretization Example} \label{sec:4:realize}
Let us discretize the QDOB for motion control of a mechanical system whose plant model is
\begin{align}
	\label{eq:}
	P_\mathrm{n}(s)=\frac{1}{Ms^2}.
\end{align}
The inverse plant model is $P_\mathrm{n}^{-1}(s)=Ms^2$.

The QDOB is discretized using two methods.
The inverse plant model $P_\mathrm{n}^{-1}(s)$ and low-pass filter $B(s)$ in \eqref{eq:B} are discretized by the backward Euler method: $s\gets (1-z^{-1})/T$ as
\begin{align}
	\label{eq:z:inv:B}
	\z[\xi_k]=\frac{M\omega_\mathrm{b} (1-z^{-1})^2}{T (1 + \omega_\mathrm{b}T - z^{-1})}\z[y_k]
\end{align}
because $B(s)P_\mathrm{n}^{-1}(s)$ is a biproper or improper-transfer function.
Subsequently, the Q-filter in \eqref{eq:Q-filter} is discretized by an exact mapping from the s-plane to the z-plane: $e^{-Ts}\gets z^{-1}$ as
\begin{subequations}
	\label{eq:}
\begin{align}
	\label{eq:z:Q}
	&\z[\hat{d}_k]=\frac{\omega_\mathrm{c} L(1+ \bar{\Psi}(z^{-1})z^{-1})}{(\omega_\mathrm{c} L+2)+(\omega_\mathrm{c} L-2) \bar{\Psi}(z^{-1})z^{-1}}\z[\xi_k-u_k]\\
	&\bar{\Psi}(z^{-1})\coloneqq z^{1-\bar{L}+N\sum_{i=0}^l\bar{U}_i}\prod_{i=0}^{l}\bar{\varphi}_i(z^{-1})\\
	&\bar{\varphi}_i(z^{-1}) \coloneqq \frac{\sum_{n=-N}^{N}w(n,N)h(n,\omega_i,U_i)z^{(n-N)\bar{U}_i}}{\sum_{n=-N}^{N}w(n,N)h(n,\omega_i,U_i)},
\end{align}
\end{subequations}
where $\bar{L}_i\coloneqq \rnd{L_i/T}$ and $\bar{U}_i\coloneqq \rnd{U_i/T}$.
This z-transform with $z$ is based on the index $k$ with the sampling time $T$ of the controller.

The discrete-time representation of \eqref{eq:z:inv:B} is obtained by using the inverse z-transform as
\begin{align}
	\label{eq:}
	\xi_k&=\frac{T\xi_{k-1} + M \omega_\mathrm{b} (y_k - 2 y_{k-1} + y_{k-2})}{T(1 + \omega_\mathrm{b}T)}.
\end{align}
The Q-filter \eqref{eq:z:Q} is transformed to collect the terms of the linear-phase low-pass filters $\bar{\Psi}(z^{-1})$ to reduce the number of buffers into
\begin{align}
	\label{eq:4:dk}
	\z[\hat{d}_k]&=\frac{\omega_\mathrm{c} L}{\omega_\mathrm{c} L+2}\z[\xi_k-u_k]+ \bar{\Psi}(z^{-1})\notag\\
	&\left(\frac{\omega_\mathrm{c} L}{\omega_\mathrm{c} L+2}\z[\xi_{k-1}-u_{k-1}]-\frac{\omega_\mathrm{c} L-2}{\omega_\mathrm{c} L+2}\z[\hat{d}_{k-1}]\right).
\end{align}
By solving \eqref{eq:4:dk} and $u_k=r_k-\mu\hat{d}_k$ with respect to $\hat{d}_k$, one obtains
\begin{subequations}
	\label{eq:}
\begin{align}
	\label{eq:}
	&\hat{d}_k=\frac{\omega_\mathrm{c} L}{(1-\mu)\omega_\mathrm{c} L+2}(\xi_k-r_k) + \mathcal{P}(\lambda_{k-1})\\
	\label{eq:calP}
	&\mathcal{P}(\lambda_{k-1})\coloneqq\mathcal{Z}^{-1}[\bar{\Psi}(z^{-1})\z[\lambda_{k-1}]]\\
	&\lambda_{k}\coloneqq\frac{\omega_\mathrm{c} L}{(1-\mu)\omega_\mathrm{c} L+2}(\xi_k-r_k) - \frac{(1-\mu)\omega_\mathrm{c} L-2}{(1-\mu)\omega_\mathrm{c} L+2}\hat{d}_k,
\end{align}
\end{subequations}
where a new parameter $\mu\in\{0,1\}$ is introduced to switch between estimation use with $\mu=0$ and compensation use with $\mu=1$.
In summary, Algorithm~\ref{alg} presents the complete discrete-time algorithm of the QDOB, including the hyperparameters, preliminary computations, functions for \eqref{eq:calP}, and real-time computation.
%##############################################################
\begin{algorithm}[t!]
\caption{Algorithm of QDOB for motion control.}
\label{alg}
{%\footnotesize
\textbf{Hyperparameters:}
\vspace{-0.5ex}
\begin{spacing}{0.4}
\begin{flalign*}
	&\mu\in\{0,1\},\
	l,
	N_\mathrm{max}\in \mathbb{Z}_{>0},\
	\omega_\mathrm{a},\
	\omega_\mathrm{b}\in \mathbb{R}_{>0},\\
	&\rho\in(0,\pi/L),\
	L,\
	M,\
	T\in \mathbb{R}_{>0}
\end{flalign*}
\end{spacing}
\vspace{-1.5ex}
\mbox{ }\\
% ---------------------------------------------------
\textbf{Preliminary computations:}
\vspace{-0.5ex}
\begin{spacing}{0.4}
\begin{flalign*}
	\hspace{1.5em}
	&\omega_\mathrm{c}=(2/L)\tan{\left(L\rho/2\right)}&\\
	&\bar{L}=\rnd{L/T}&\\
	&c=(1/2)(T\omega_\mathrm{a}/\pi)^{1/l}&\\
	&\bm{\mathrm{for}}\ i=1\ \ldots\ l\ \bm{\mathrm{do}}&\\
	&\hspace{1.5em}\bm{\mathrm{if}}\ i==1\ \bm{\mathrm{then}}&\\
	&\hspace{1.5em}\hspace{1.5em} U_{i}=T&\\
	&\hspace{1.5em}\bm{\mathrm{else}}&\\
	&\hspace{1.5em}\hspace{1.5em} U_{i}=\pi/\omega_{i-1}&\\
	&\hspace{1.5em}\bm{\mathrm{end\ if}}&\\
	&\hspace{1.5em}\bar{U}_{i}=\rnd{U_i/T}&\\
	&\hspace{1.5em}\omega_i={2\pi c}/{U_i}&\\
	&\bm{\mathrm{end\ for}}&\\
	&N=\min\{\mathrm{floor}((\bar{L}-1)/\textstyle\sum_{i=1}^l\bar{U}_i),\ N_\mathrm{max}\}&\\
	&\eta=\bar{L}-N\textstyle\sum_{i=1}^l\bar{U}_i&
\end{flalign*}
\end{spacing}
\hrulefill\\
% ---------------------------------------------------
\textbf{Function $\mathcal{P}$:}\\
\hspace{1.5em}\textbf{Input:} $\lambda_{k-1}$,\ \ \textbf{Output:} $\lss{l}\theta_k$
\vspace{-0.5ex}
\begin{spacing}{0.4}
\begin{flalign*}
	\hspace{1.5em}&\lss{0}\theta_k=\lambda_{k-\eta}&\\
	&\bm{\mathrm{for}}\ i=1\ \ldots\ l\ \bm{\mathrm{do}}&\\
	&\hspace{1.5em}\bm{\mathrm{for}}\ n=-N\ \ldots\ N\ \bm{\mathrm{do}}&\\
	&\hspace{1.5em}\hspace{1.5em}\lss{i}\theta_k\mathrel{+}=w(n,N)h(n,\omega_i,U_i)\ \lss{i-1}\theta_{k+(n-N)\bar{U}_i}&\\
	&\hspace{1.5em}\hspace{1.5em}\lss{i}\gamma_k\mathrel{+}=w(n,N)h(n,\omega_i,U_i)&\\
	&\hspace{1.5em}\bm{\mathrm{end\ for}}&\\
	&\hspace{1.5em}\lss{i}\theta_k=\lss{i}\theta_k/\lss{i}\gamma_k&\\
	&\bm{\mathrm{end\ for}}&
\end{flalign*}
\end{spacing}
\vspace{1ex}
\hspace{1.5em}\textbf{Sub-functions:}
\vspace{-1ex}
\begin{spacing}{0.4}
\begin{flalign*}
	\hspace{1.5em}&h(n,\omega_i,U_i)=\left\{
	\begin{array}{cl}
		{U_i\omega_i}/{\pi}&\mathrm{if}\ n=0\\
		{\sin(nU_i\omega_i)}/{(n\pi)}&\mathrm{if}\ n\neq0\\
	\end{array}
	\right.&\\
	&w(n,N)=0.42+0.5\cos({n\pi}{/N})+0.08\cos({2n\pi}/{N})&
\end{flalign*}
\end{spacing}
\hrulefill\\
% ---------------------------------------------------
\textbf{Real-time computation}\\
\hspace{1.5em}\textbf{Inputs:} Reference ${r}_k$,\ Response $y_k$\\
\hspace{1.5em}\textbf{Output:} Control input $u_k$,\ Estimated disturbance $\hat{d}_k$
\vspace{-0.5ex}
\begin{spacing}{0.4}
\begin{flalign*}
	\hspace{1.5em}
	&\xi_k=\frac{1}{T(1 + \omega_\mathrm{b}T)}[T\xi_{k-1} + M \omega_\mathrm{b} (y_k - 2 y_{k-1} + y_{k-2})]&\\
	&\hat{d}_k=\frac{\omega_\mathrm{c} L}{(1-\mu)\omega_\mathrm{c} L+2}(\xi_k-r_k) + \mathcal{P}(\lambda_{k-1})&\\
	&\lambda_k=\frac{\omega_\mathrm{c} L}{(1-\mu)\omega_\mathrm{c} L+2}(\xi_k-r_k) - \frac{(1-\mu)\omega_\mathrm{c} L-2}{(1-\mu)\omega_\mathrm{c} L+2}\hat{d}_k&\\
	&u_k=r_k-\mu\hat{d}_k&
\end{flalign*}
\end{spacing}
\vspace{2ex}
}
\end{algorithm}
%##############################################################

\subsection{Guide to Hyperparameter Tuning} \label{sec:4:hyp}
Algorithm~\ref{alg} has nine hyperparameters: $\mu$, $l$, $N_\mathrm{max}$, $\omega_\mathrm{a}$, $\omega_\mathrm{b}$, $\rho$, $L$, $M$, and $T$ to be tuned.

The parameter $\mu\in\{0,1\}$ is set to $\mu=0$ when the QDOB is only used to estimate a quasiperiodic disturbance and to $\mu=1$ when the QDOB is used to compensate for a quasiperiodic disturbance.
For example, the estimation with $\mu=0$ helps to verify the implemented algorithm for the QDOB before applying compensation $\mu=1$.

The number of stages $l\in \mathbb{Z}_{>0}$ and maximum order $N_\mathrm{max}\in \mathbb{Z}_{>0}$ are the parameters for the linear-phase low-pass filter $\Phi(s)$ described in Section~\ref{sec:LP-LPF}.
The parameters $l$ and $N_\mathrm{max}$ are determined by considering the trade-off, in which an increase in $l$ and/or $N_\mathrm{max}$ improves the filter $\Phi(s)$ to be closer to an ideal linear-phase low-pass filter but increases the computational cost.
Empirically, a single-digit integer for $l$ and three-digit integer for $N_\mathrm{max}$ are sufficiently practical.
The frequency response of the filter $\Phi(s)$ helps evaluate if $l$ and $N_\mathrm{max}$ are large enough to approximate an ideal linear-phase low-pass filter.

The angular frequencies $\omega_\mathrm{a}$, $\omega_\mathrm{b}$, and $\rho$ should be tuned according to the following trade-off.
The cutoff frequencies $\omega_\mathrm{a}$ for \eqref{eq:wi} and $\omega_\mathrm{b}$ for \eqref{eq:B} need to satisfy $\omega_\mathrm{a}\ll\omega_\mathrm{b}$, which leads to the phase of the open-loop transfer function \eqref{eq:Gamma} within $-90$ deg. to $90$ deg. at frequencies below the Nyquist frequency (Section~\ref{sec:3:ns}).
This would enable the avoidance of the Bode's sensitivity integral at frequencies less than the Nyquist frequency, which results in the non-amplification of aperiodic disturbances and the proper harmonic suppression frequencies (Section~\ref{sec:3:ST}).
Moreover, the phase between $-90$ deg. to $90$ deg. has wide stability margin (Sections~\ref{sec:3:ns} and \ref{sec:3:rs}).
Under the condition $\omega_\mathrm{a}\ll\omega_\mathrm{b}$, the cutoff frequency $\omega_\mathrm{a}$ needs to be higher than target harmonic frequencies, as $\omega_0,\ 2\omega_0,\ 3\omega_0,\ \ldots<\omega_\mathrm{a}\ll\omega_\mathrm{b}$.
Additionally, the separation frequency $\rho\in(0,\pi/L)$ needs to be high to extend the harmonic suppression bandwidth.
Conversely, $\omega_\mathrm{b}$ and $\rho$ need to be low enough to satisfy the robust stability condition \eqref{eq:robust:condition} (Section~\ref{sec:3:rs}).

The values of $L$, $M$, and $T\in \mathbb{R}_{>0}$ are the identified parameters from a target quasiperiodic disturbance, mass of the supposed plant $1/(Ms^2)$, and sampling time of the controller, respectively.

\section{Experiments} \label{sec:5}
\subsection{Frequency Response} \label{sec:5a}
\subsubsection{Setup}
\begin{figure}[t!]
	\begin{center}
		\includegraphics[width=0.65\hsize]{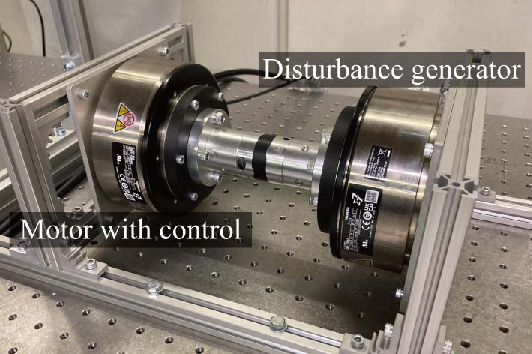}\\
		(a)\\
		\vspace{1ex}
		\includegraphics[width=0.9\hsize]{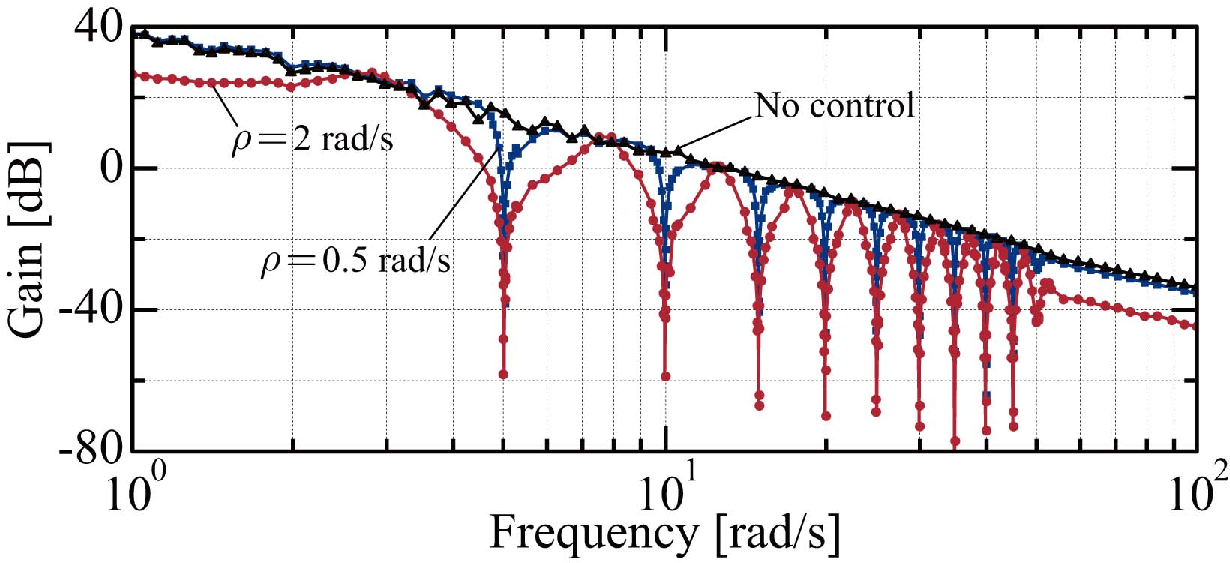}\\
		(b)\\
		\vspace{1ex}
		\includegraphics[width=0.9\hsize]{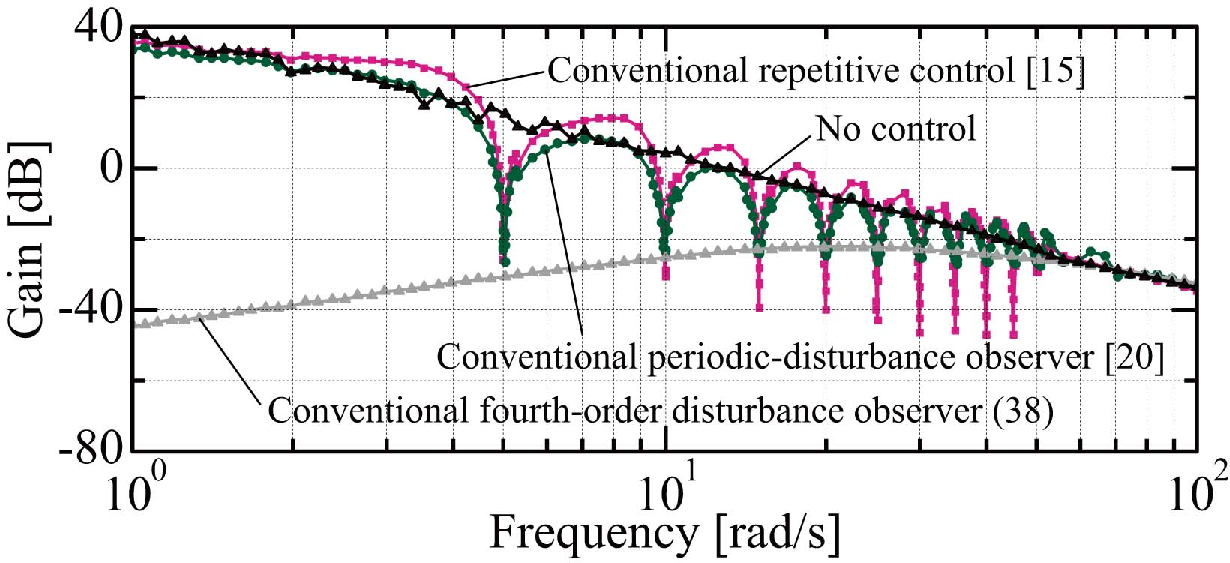}\\
		(c)
	\end{center}
	\vspace{-2ex}
	\caption{Frequency-response experiment. (a) Setup. (b) Results of the QDOB. (c) Results of the conventional methods \cite{2014_Chen_DOBbasedRC}, \cite{2018_Muramatsu_APDOB}, and \eqref{eq:exp:highDOB}.}\label{fig:FR}
\end{figure}

The frequency response of the QDOB and plant from the external torque $v$ to the angle $y$ was validated and compared with those of conventional repetitive control \cite{2014_Chen_DOBbasedRC}, periodic-disturbance observer \cite{2018_Muramatsu_APDOB}, and fourth-order disturbance observer:
\begin{subequations}
	\label{eq:exp:highDOB}
\begin{align}
	u(t)&=r(t)-\hat{d}(t)\\
	\mathcal{L}[\hat{d}(t)]&=Q(s)(P_\mathrm{n}^{-1}(s)\mathcal{L}[y(t)]-\mathcal{L}[u(t)])\\
	Q(s)&=\frac{c_{2}s^{2}+c_{1}s^{1}+c_{0}}{(s+g)^4},\
	c_i=\frac{4!}{(4-i)!i!}g^{4-i}.
\end{align}
\end{subequations}
In this experiment, two direct-drive motors (SGMCS-02BDC41 from YASKAWA Electric Corporation; moment of inertia: $28.0\times10^{-4}\ \mathrm{kg\!\cdot\!m^2}$) connected by a coupling were used (Fig.~\ref{fig:FR}(a)), where the direct-drive mechanism reduced unintended disturbances.
The left motor was controlled using one of the proposed or conventional methods without any other controllers.
The right motor generated external torque $v=a_\mathrm{v}\sin(\omega_\mathrm{v}t)$.
For all conditions, the frequency responses were measured with the amplitude $a_\mathrm{v}=0.3$ Nm at the angular frequencies $\omega_\mathrm{v}=10^{0.025i}$ rad/s with $i=0,\ \ldots,\ 80$.
For the proposed QDOB, conventional repetitive control, and periodic-disturbance observer, the frequency responses were additionally measured with the amplitude $a_\mathrm{v}=2$ Nm at the angular frequencies around the harmonic frequencies, such that $\omega_\mathrm{v}=10^{\log_{10}(5j)+0.005i-0.02}$ with $i=0,\ \ldots,\ 8$ and $j=1,\ \ldots,\ 10$ or $\omega_\mathrm{v}=10^{\log_{10}(5j)+0.001i-0.002}$ with $i=0,\ \ldots,\ 4$ and $j=1,\ \ldots,\ 10$.
Each sinusoidal response was tested for 40 or 60 s, and the discrete Fourier transform was applied to the steady-state response, eliminating the initial 20-s transient response to compute the gain.
As the QDOB, Algorithm~\ref{alg} derived in Section~\ref{sec:4} was implemented with the parameters $l=3$, $N_\mathrm{max}=256$, $\omega_\mathrm{a}=50$ rad/s, $\omega_\mathrm{b}=100$ rad/s, $\rho\in\{0.5,\ 2\}$ rad/s, $M=56.13\times10^{-4}$ $\mathrm{kg\!\cdot\!m^2}$, $L=2\pi/5$ s, and $T=2\times10^{-4}$ s.
The conventional repetitive control used the same zero-phase low-pass filter as that for the QDOB with $\alpha=0.9$.
In the conventional periodic-disturbance observer, its Q-filter used $\gamma=0.5$ and a cutoff frequency of $g=50$ rad/s, and the pseudo-differentiation used a cutoff frequency of 100 rad/s.
The conventional fourth-order disturbance observer used a cutoff frequency of $g=50$ rad/s.

%##############################################################
%##############################################################
\subsubsection{Results}
Fig.~\ref{fig:FR}(b) depicts three frequency responses: with no control, with the QDOB using $\rho=0.5$ rad/s, and with the QDOB using $\rho=2$ rad/s.
The frequency response with no control obeyed the gain of $1/(Ms^2)$.
This indicates that the model $1/(Ms^2)$ of the plant was accurate, and disturbances not generated by the right motor were negligible.
The effect of the implemented control appeared as the differences from the frequency response without control.
The frequency response with the QDOB realized sharp band-stop frequencies at the harmonic frequencies from the first harmonic at 5 rad/s to the tenth harmonic at 50 rad/s.
The target harmonic frequencies were determined by the cutoff frequency $\omega_\mathrm{a}$ of 50 rad/s.
Compared to the separation frequency $\rho=0.5$ rad/s, $\rho=2$ rad/s extended the suppression bandwidth around the harmonic frequencies without amplification of aperiodic disturbances and deviation of the harmonic suppression frequencies.
From 60 to 100 rad/s, the gain decreased as the separation frequency $\rho$ increased according to \eqref{eq:ST:middle}.

The proposed QDOB was compared with the conventional methods in Fig.~\ref{fig:FR}(c).
The conventional repetitive control \cite{2014_Chen_DOBbasedRC} showed aperiodic disturbance amplification due to the trade-off between the wideband suppression and the amplification.
The conventional periodic-disturbance observer \cite{2018_Muramatsu_APDOB} showed wider suppression around the harmonics and non-amplification of the aperiodic disturbances from 1 to 35 rad/s.
However, the harmonic suppression performance was less than that of the repetitive control.
The amplification appeared from 35 rad/s, and the high-order harmonic suppression frequencies deviated slightly.
Compared to them, the QDOB achieved lower gain at the harmonic frequencies (5, 10, 15, $\ldots$, 45 rad/s), wideband suppression around the harmonic frequencies with $\rho=2$ rad/s, non-amplification around the aperiodic-disturbance frequencies (2.5, 7.5, 12.5, $\ldots$, 47.5 rad/s), and non-deviation of the harmonic suppression frequencies, as shown in Fig.~\ref{fig:FR}(b).
The lower gain of the QDOB at the harmonic frequencies was caused by the zero-phase low-pass filter integrated with each time delay.
Compared to the conventional fourth-order disturbance observer, the QDOB showed the less gain at the harmonic frequencies.
The gain of the fourth-order disturbance observer was less than that of the QDOB at aperiodic-disturbance frequencies (around 7.5, 12.5, 17.5, $\ldots$, 47.5 rad/s), which was an intended result because the QDOB was not supposed to suppress aperiodic disturbances.

%##############################################################
\begin{figure}[t!]
	\begin{center}
		\includegraphics[width=0.65\hsize]{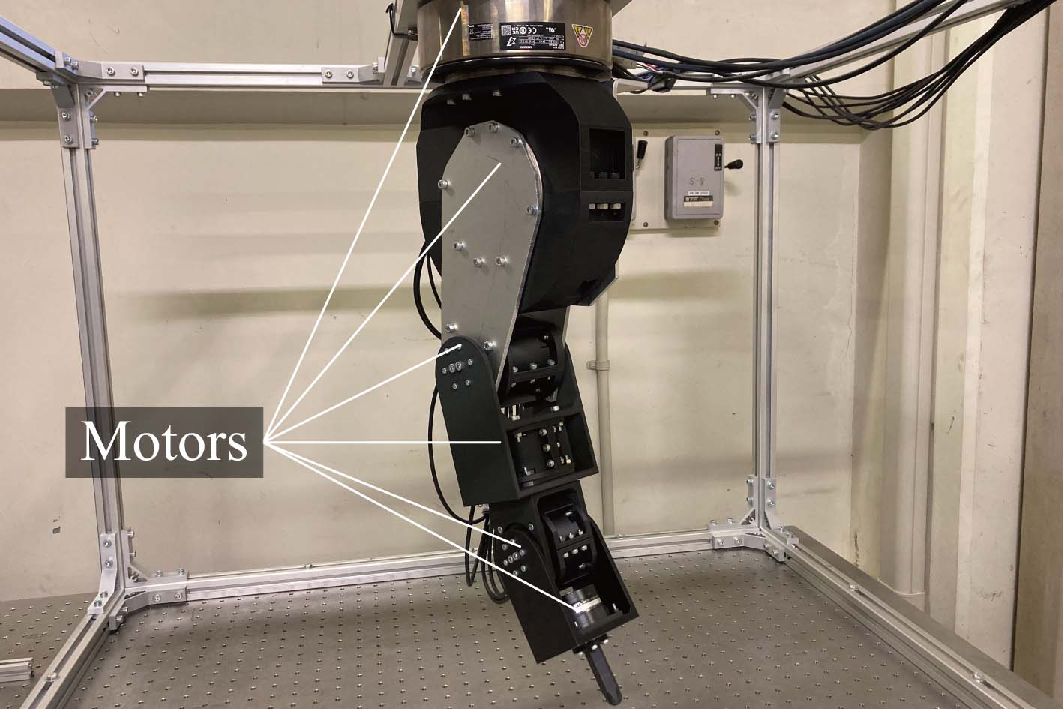}\\
		(a)\\
		\vspace{1ex}
		\includegraphics[width=0.9\hsize]{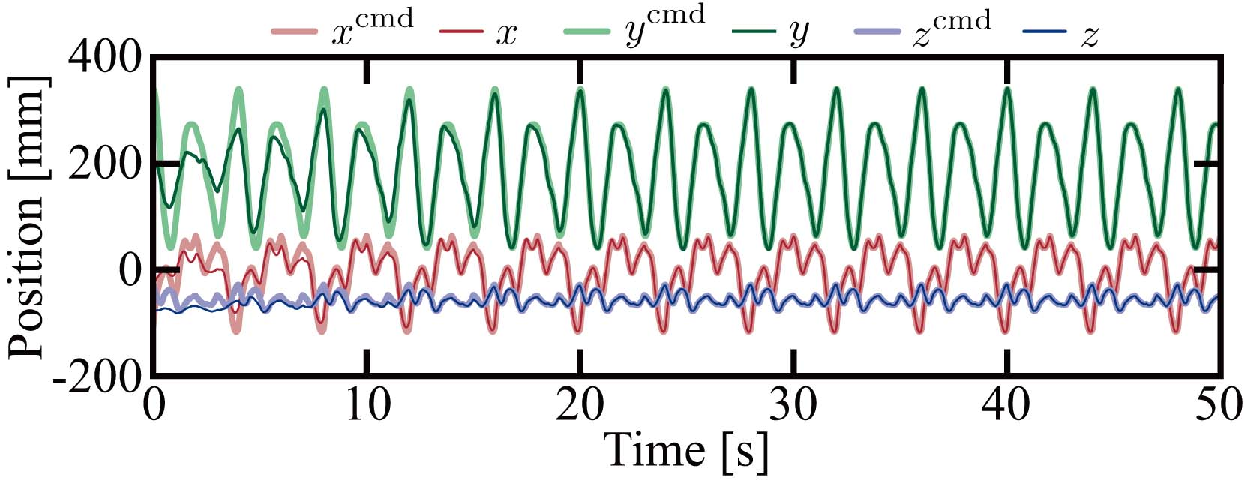}\\
		(b)\\
		\includegraphics[width=0.9\hsize]{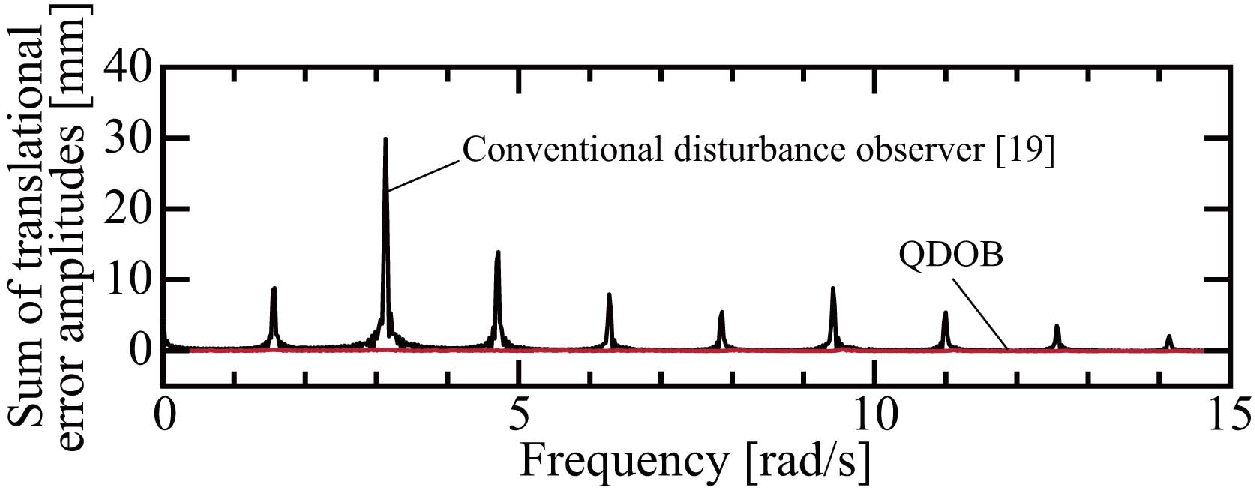}\\
		(c)\\
		\vspace{1ex}
		\includegraphics[width=0.9\hsize]{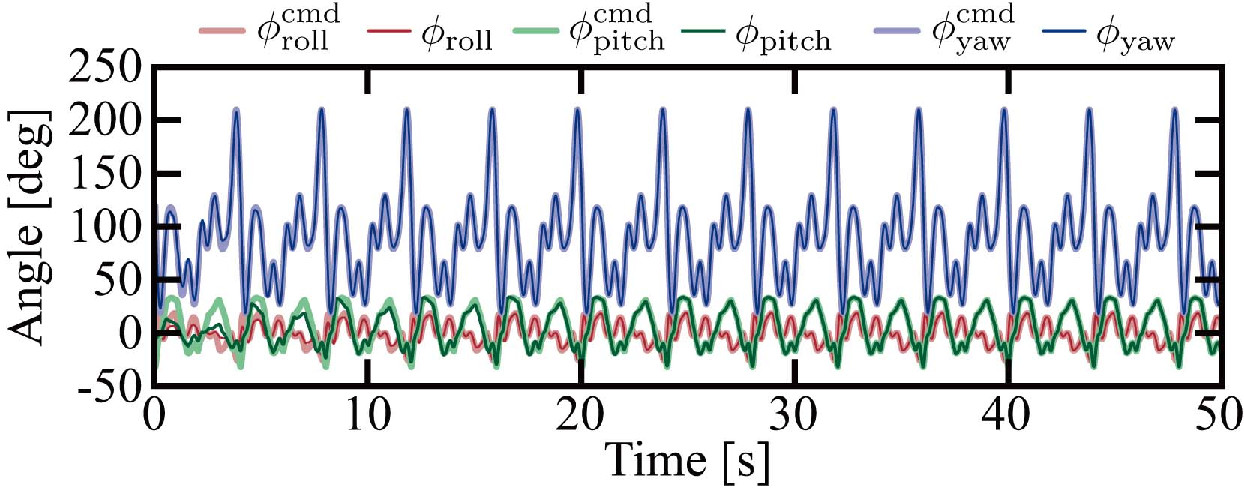}\\
		(d)\\
		\includegraphics[width=0.9\hsize]{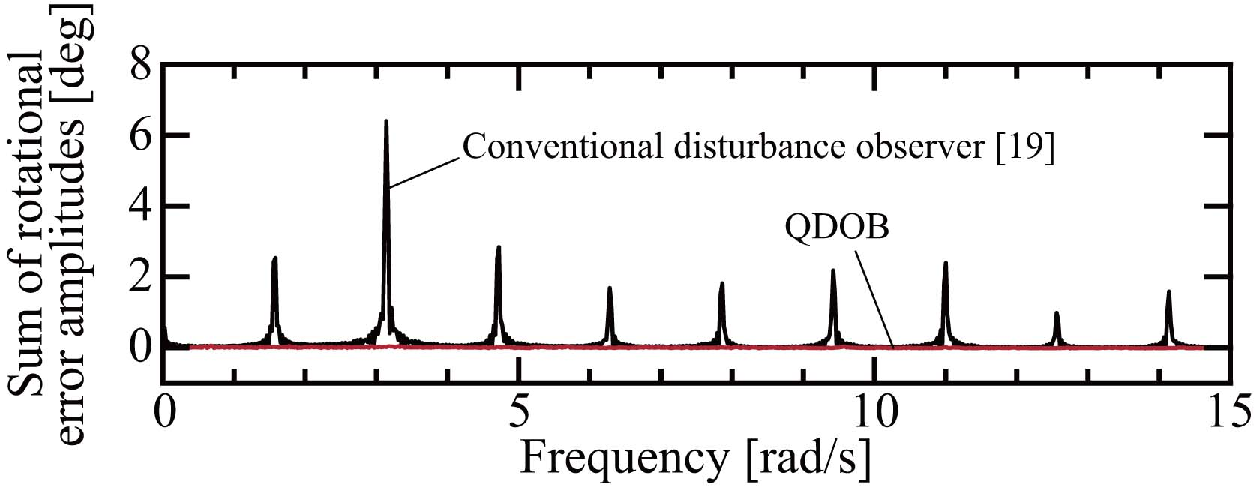}\\
		(e)
	\end{center}
	\vspace{-2ex}
	\caption{Manipulator-control experiment. (a) Setup. (b) Position (x-y-z) results of the QDOB. (c) Sum of the amplitudes of the discrete Fourier transform translational errors in the steady state. (d) Orientation (roll-pitch-yaw) results of the QDOB. (e) Sum of the amplitudes of the discrete Fourier transform rotational errors in the steady state.}\label{fig:SR}
\end{figure}
%##############################################################
\subsection{Manipulator Control} \label{sec:5b}
\subsubsection{Setup}
The QDOB was used to control a six-degree-of-freedom manipulator (SGM7E-04CFA41 from YASKAWA Electric Corporation; MDH-7018-648KE and MDH-6012-500KE from Microtech Laboratory Inc.) with proportional-and-derivative angle control and feedforward control in the joint space (Fig.~\ref{fig:SR}(a)).
The moments of inertia of SGM7E-04CFA41, MDH-7018-648KE, and MDH-6012-500KE were $77.0\times10^{-4}$, $0.99\times10^{-4}$, and $0.42\times10^{-4}$ $\mathrm{kg\!\cdot\!m^2}$, respectively.
The effect of the QDOB was verified and compared with that of a conventional disturbance observer \cite{2015_Sariyildiz_DOB}.
The outer joint-space proportional-and-derivative controller was
\begin{subequations}
	\label{eq:}
\begin{align}
	\label{eq:}
	\dot{\hat{\bm{e}}}(t)&=\bm{G}(\bm{e}(t)-\hat{\bm{e}}(t)),\ \bm{e}(t)=\bm{\theta}^\mathrm{cmd}(t)-\bm{\theta}(t)\\
	\ddot{\hat{\bm{\theta}}}^\mathrm{cmd}(t)&=\bm{G}^2\bm{\theta}^\mathrm{cmd}(t)-2\bm{G} \dot{\hat{\bm{\theta}}}^\mathrm{cmd}(t) - \bm{G}^2\hat{\bm{\theta}}^\mathrm{cmd}(t)\\
	\bm{r}(t)&=\bm{K}_\mathrm{p}\bm{e}(t) + \bm{K}_\mathrm{d} \dot{\hat{\bm{e}}}(t) + \bm{M}\ddot{\hat{\bm{\theta}}}^\mathrm{cmd}(t),
\end{align}
\end{subequations}
where the derivative gain, proportional gain, moment of inertia matrix, and cutoff frequencies for the pseudo-differentiation were set as
$\bm{K}_\mathrm{d}=\bm{\mathrm{diag}}$(0.5, 0.5, 0.4, 0.1, 0.05, 0.05) $\mathrm{N\!\cdot\!m\!\cdot\!s}/\mathrm{rad}$,
$\bm{K}_\mathrm{p}=\bm{\mathrm{diag}}$(3, 3, 1.5, 0.5, 0.2, 0.2) $\mathrm{N\!\cdot\!m}/\mathrm{rad}$,
$\bm{M}=\bm{\mathrm{diag}}$(8, 30, 20, 0.5, 1, 0.1)$\times10^{-3}$ $\mathrm{kg\!\cdot\!m^2}$, and
$\bm{G}=\bm{\mathrm{diag}}$(200, 200, 200, 200, 200, 100) rad/s, respectively.
The variables $\bm{\theta}^\mathrm{cmd}(t)\in \mathbb{R}^6$, $\bm{\theta}(t)\in \mathbb{R}^6$, and $\bm{e}(t)\in \mathbb{R}^6$ denote the command angle, response angle, and angle error, respectively.
The QDOB used the same algorithm (Algorithm~\ref{alg}) and parameters as those in Section~\ref{sec:5a}, except for the parameters: moment of inertia $\bm{M}$, separation frequency $\rho=2$ rad/s, period $L=4$ s, and sampling time $T=2\times10^{-3}$ s.
The conventional disturbance observer used a cutoff frequency of 50 rad/s for its Q-filter.
The manipulator was position-controlled with periodic position and orientation commands for the end-effector with a period of four seconds, where quasiperiodic disturbances such as gravity and friction occurred.
The periodicity was caused by the periodic commands and was quasi as no actual disturbances can satisfy $d(t)=d(t-L)$ strictly, although the degree of quasiperiodicity was small in this case.
Note that the QDOB does not require identification of phenomena and models for quasiperiodic disturbances, but it requires a period of quasiperiodicity while being robust against errors in the identified period.

\subsubsection{Results}
The command and response waveforms from 0 s to 50 s of the end-effector position (x-y-z) and orientation (roll-pitch-yaw) with the QDOB are shown in Figs.~\ref{fig:SR}(b) and (d), respectively.
They show that the QDOB needed almost four cycles (16 s) of the 4-s period to be effective, owing to the buffer implementing the linear-phase low-pass filter.
The discrete Fourier transform was applied to the steady-state errors of the position and orientation from 30 s to 180 s for both the QDOB and conventional disturbance observer \cite{2015_Sariyildiz_DOB}.
Figs.~\ref{fig:SR}(c) and (e) show the sum of the amplitudes of the Fourier-transform position and orientation errors, respectively.
The harmonic suppression of the QDOB from the first harmonic (1.57 rad/s) to the ninth harmonic (14.14 rad/s) was observed for both position and orientation, compared to the conventional disturbance observer.

\section{Conclusion} \label{sec:6}
This paper proposed the QDOB to estimate and compensate for quasiperiodic disturbances.
The QDOB is expected to improve control accuracy of practical automatic control systems suffering harmonics.
Actual periodic disturbances composed of harmonics usually become quasiperiodic because of perturbations in each cycle, identification errors of the period, variations in the period, and/or aperiodic disturbances.
The QDOB can suppress harmonics with the robustness against the quasiperiodicity (the wideband harmonic suppression) without amplifying aperiodic disturbances and deviation of harmonic suppression frequencies, unlike conventional repetitive control and periodic disturbance observers~\cite{2002_Steinbuch_RC,2007_Steinbuch_RC,2008_PipeleersRC,2014_Chen_DOBbasedRC,2021_Nie_DOBbasedRC,2018_Muramatsu_APDOB,2019_Muramatsu_EnPDOB,2023_Tanaka_PDOB,2023_Yang_PDOB,2021_Lai_PDOB,2023_Li_PDOB}.
Furthermore, its stability can be nominally guaranteed and is robust against modeling errors.
These characteristics were validated through the experiments with motors and manipulator.
The QDOB is applicable to plants such that \eqref{eq:plant} and \eqref{eq:Pdelta} covering a wide range of practical applications, including the experimental setups.
However, there is room for improvement in further extending the applicable plants, such as unstable, non-minimum phase, multi-input-multi-output, and nonlinear plants.


%==========================================================================
\vspace{-7mm}
\begin{IEEEbiography}[{\includegraphics[width=1in,height=1.25in,clip,keepaspectratio]{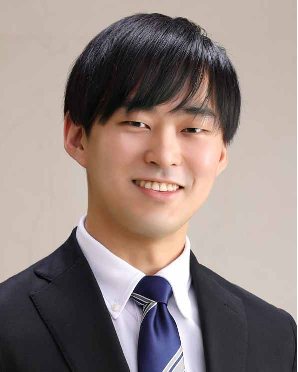}}]{Hisayoshi Muramatsu}
received the B.E. degree in system design engineering and the M.E. and Ph.D. degrees in integrated design engineering from Keio University, Yokohama, Japan, in 2016, 2017, and 2020, respectively.
From 2020 to 2024, he was an Assistant Professor with the Mechanical Engineering Program, Hiroshima University, Higashihiroshima, Japan, where he is currently an Associate Professor.
His research interests include motion control, periodic/aperiodic, harmonics, mobile robots, and human-robot interaction.
\end{IEEEbiography}

\vfill
%==========================================================================
\end{document}